# Super-accuracy calculation for the width of a Voigt profile

Yihong Wang, Bin Zhou[*], Rong Zhao, Bubin Wang

School of Energy and Environment, Southeast University, Nanjing 210096, China

[*]zhoubinde@seu.edu.cn

**Abstract** A simple approximation scheme to describe the width of the Voigt profile as a function of the relative contributions of Gaussian and Lorentzian broadening is presented. The proposed approximation scheme is highly accurate and provides accuracy better than $10^{-17}$ for arbitrary $\alpha_L/\alpha_G$ ratios. In particular, the accuracy reaches an astonishing $10^{-34}$ (quadruple precision) in the domain $0 \leq \alpha_L/\alpha_G \leq 0.2371 \cup \alpha_L/\alpha_G \geq 33.8786$.

1. **Introduction**

The Voigt profile is the convolution of the Gaussian and Lorentzian profiles, which is widely used in optics, laser physics, plasma and spectroscopy[1-6]. In actual application, the half width at half maximum of the Voigt profile, $\alpha_V$, is important[3-6]. Unfortunately, there is no analytically exact expression to describe the width of the Voigt profile as a function of the widths of the Lorentzian and Gaussian profiles, $\alpha_L$ and $\alpha_G$, respectively and thus many approximations have been presented in the past to find simple relationships, i.e. composed of basic elementary functions only, between $\alpha_V$, $\alpha_L$ and $\alpha_G$[7-11].

As far as we know, the approximation expression developed by Olivero and Longbothum (Eq.5. in [11]) is the most accurate, and its estimation error is less than 0.01%. Although this approximation is sufficient for the most practical tasks, the more accurate approximation may also be required. Several highly accurate codes such as ACM Algorithm 680[12, 13] (14 significant digits stated accuracy) and ACM Algorithm 916[14] (20 significant digits) or arbitrary precision codes[15, 16] are indispensable for evaluates the Voigt profile, but not necessarily fast. However, in some cases where we only focus on broadening[3-6], it is unwise to use these highly accurate codes. Therefore, a highly accurate approximate scheme specially used to evaluate the half width at half maximum of the Voigt profile is urgently needed.

In this work we present a simple approximation scheme to describe the width of the Voigt profile as a function of the relative contributions of Gaussian and Lorentzian broadening. The numerical calculations suggest that the proposed approximation expression can achieve super-accuracy calculation for Voigt profiles for arbitrary $\alpha_L/\alpha_G$ ratios.

2. **Methodology and derivation**

It is convenient to introduce the Voigt function when considering the width of the Voigt profile. The Voigt function (normalized to $\sqrt{\pi}$) is defined by

$$K(x,y) = \frac{y}{\pi}\int_{-\infty}^{+\infty} \frac{e^{-t^2}}{(x-t)^2+y^2}dt \qquad (1)$$
$$= \text{Re}[e^{-z^2}\text{erfc}(-iz)],$$

where the dimensionless variables $x = \sqrt{\ln 2}(v-v_0)/\alpha_G$, $y = \sqrt{\ln 2}\alpha_L/\alpha_G$ are a measure for the distance from the peak center $v_0$ and for the ratio of Lorentzian to Gaussian width, respectively, and $z = x+iy$. At the line center $x = 0$ the Voigt function can be expressed as the exponentially scaled complementary error function $K(0,y) = \exp(y^2)\text{erfc}(y)$[17]. Thus, the half width scale in units of Gaussian broadening of a Voigt profile, $\Gamma$, for a given value of $y$ is given by the following equation:

$$K(\Gamma, y) = \frac{1}{2}e^{y^2}\text{erfc}(y), \qquad (2)$$

where $\Gamma$ is defined as $\Gamma = \sqrt{\ln 2}\alpha_V/\alpha_G$. The variables $x$ and $y$ are more convenient for lines which are predominantly Gaussian while the variables $X=x/y$ and $\eta=1/y$ are more convenient for lines which are predominantly Lorentz [18]. Thus, the half width scale in units of Lorentzian broadening of a Voigt profile, $\Gamma'$, for a given value of $\eta$ is given by the following equation:

$$K(\Gamma'/\eta, 1/\eta) = \frac{1}{2} e^{1/\eta^2} \text{erfc}(1/\eta), \qquad (3)$$

where $\Gamma'$ is defined as $\Gamma' = \alpha_V/\alpha_L$. It is obvious that $\Gamma = \sqrt{\ln 2}$ at $y = 0$ and $\Gamma' = 1$ at $\eta = 0$ from the definition of Voigt function.

2.1 The width of Voigt profile with small $y$ ($0 \le y \le 0.6993$)

The following infinite series are given to calculate the width of Voigt profile with small $y$:

$$\Gamma(y) = \sum_{n=0}^{\infty} p_n y^n, \qquad (4)$$

and simultaneous Eq. 1, Eq. 2 shows that $\Gamma$ satisfies the following equation:

$$e^{-\Gamma^2} \sin(2\Gamma y) \text{Im}[\text{erfc}(y - i\Gamma)] + e^{-\Gamma^2} \cos(2\Gamma y) \text{Re}[\text{erfc}(y - i\Gamma)] = \frac{1}{2} \text{erfc}(y). \qquad (5)$$

Using the series analysis method for Eq. 5, the recurrence equation satisfied by the expression of the $(r+1)$-th coefficient $p_r$ ($r >= 1$) of each order in Eq. 4 can be obtained as follows:

$$p_r = \frac{e^{p_0^2}}{-2p_0} [s_r - q_r \exp(-p_0^2) - \sum_{\substack{j+k+m=r \\ 0 \le j \le r-1, 0 \le k \le r-1, 1 \le m \le r}} f_j g_m l_k - \sum_{j=1}^{r} \sum_{k=0}^{r-j} q_{r-j-k} f_k h_j$$
$$- \sum_{k=1}^{r-1} q_{r-k} f_k - \sum_{n=2}^{r} \frac{\alpha_{n,r,r} b_n}{n!}], \qquad (6)$$

where each coefficient in the above expression can be calculated by the first $r$ coefficients, $p_0, p_1, \ldots, p_{r-1}$. The derivation of Eq. 6 and the definition of each coefficient can be found in Appendix A. We notice the fact that $\Gamma = \sqrt{\ln 2}$ at $y = 0$, which indicates that the first coefficient $p_0 = \sqrt{\ln 2}$. Therefore, the coefficient of each order in Eq. 4 can be obtained immediately by solving the recurrence equation. For example, the first four coefficients are shown as follows:

$$p_0 = \sqrt{\ln 2},$$
$$p_1 = \text{erfi}(\sqrt{\ln 2}) - \frac{1}{\sqrt{\pi \ln 2}},$$
$$p_2 = -\frac{(2\theta_1^2 - 1)\theta_3^2}{2\theta_1} - \frac{2\theta_1^4 \theta_2^2 + 6\theta_1^2 + 1}{2\theta_1^3 \theta_2^2} + \frac{4\theta_3}{\theta_2}, \qquad (7)$$
$$p_3 = \frac{4(4\theta_1^2 - 1)\theta_3}{\theta_1^2 \theta_2^2} + \frac{2}{3}(2\theta_1^2 - 3)\theta_3(\theta_3^2 + 1) - \frac{\theta_1^2(56\theta_1^2 + 6) + 3}{6\theta_1^5 \theta_2^3} - \frac{16\theta_1^4(3\theta_3^2 + 1) - 2\theta_1^2(21\theta_3^2 + 4) - 3\theta_3^2}{6\theta_1^3 \theta_2},$$

where constants $\theta_1 = \sqrt{\ln 2}$, $\theta_2 = \sqrt{\pi}$ and $\theta_3 = \text{erfi}(\sqrt{\ln 2})$, respectively and erfi(.) is the complementary error function. In addition, the first 31 terms with 32 significant digits are given in the Table A.1.

2.2 The width of Voigt profile with large $y$ ($y \ge 8.2507$)

Numerical analysis suggests that it is convenient to analyze the relationship between $\Gamma'^2$ and $\eta$, and a simple series is given as following (Appendix B):

$$\Gamma'^2(\eta) = 1 + \frac{3}{2}\eta^2 - \frac{3}{2^2}\eta^4 + \frac{15}{2^3}\eta^6 - \frac{243}{2^5}\eta^8 + \frac{2493}{2^6}\eta^{10} - \frac{927}{2^2}\eta^{12} + \frac{405783}{2^8}\eta^{14} - \frac{25390179}{2^{11}}\eta^{16}$$
$$+ \frac{446848569}{2^{12}}\eta^{18} - \frac{1089694161}{2^{10}}\eta^{20} + \frac{46704949839}{2^{12}}\eta^{22} - \frac{8735832539883}{2^{16}}\eta^{24} \qquad (8)$$
$$+ \frac{221377058104455}{2^{17}}\eta^{26} - \frac{6044700753428715}{2^{18}}\eta^{28} + \frac{353911508743725891}{2^{20}}\eta^{30} - \ldots$$

In the above expression, the coefficients of odd terms are all zero, and the even terms are all rational fractions, which is conducive to high precision calculation.

2.3 The width of Voigt profile with middle y (0.6993 < y < 8.2507)

Because of the best uniform approximation property of Chebyshev polynomial, the best uniform approximation expression of the Voigt profile with middle y (0.6993 < y < 8.2507) can be got. Numerical analysis suggests that it is useful to introduce the non-dimensional quantities $R = \alpha_V/(\alpha_L + \alpha_G)$ and $D = (\alpha_L - \alpha_G)/(\alpha_L + \alpha_G) = (y - \sqrt{\ln 2})/(y + \sqrt{\ln 2})$, and the best uniform approximation expression is given as following:

$$R(D) = \sum_{k=0}^{N} R(D_k) \prod_{\substack{0 \le m \le N \\ m \ne k}} \frac{D - D_m}{D_k - D_m}, \qquad (9)$$

where $(D_k, R(D_k))$ are a set of $N$ sample points without two identical points of $D_k$, and $D_k$ are given as following:

$$D_k = 0.3648 + 0.4518 \cos\left(\frac{2k-1}{2n+2}\pi\right), k = 1, 2, \cdots, N+1. \qquad (10)$$

For computational convenience, the above mathematical expression can be changed into the following equivalent series:

$$R(D) = \sum_{k=0}^{N} u_k D^k, \qquad (11)$$

where the coefficients $u_k$ can be calculated by numerical method, and the first 31 coefficients with 32 significant digits are given in the Table C.1.

Thus, the computation flow for the basic and the main approximations can be maintained as

$$\alpha_V \approx \begin{cases} \alpha_G \Gamma(y)/\sqrt{\ln 2}, & \text{Eq.(4) for } 0 \le y \le 0.6993 \\ (\alpha_L + \alpha_G) R(D), & \text{Eq.(11) for } 0.6993 < y < 8.2507 \\ \alpha_L \Gamma'(\eta), & \text{Eq.(8) for } y \ge 8.2507. \end{cases} \qquad (12)$$

## 3. Error analysis

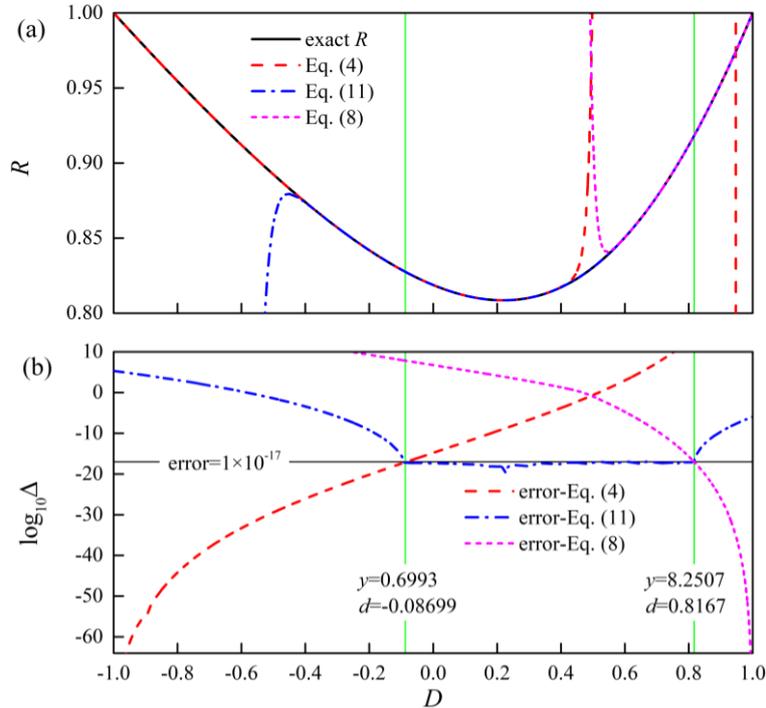

Fig. 1(a) The comparison of each width functions and (b) the logarithm of relative error for the width of Voigt profile, respectively, in the $R$, $D$ format

Define the relative errors for the width of Voigt profile in form

$$\Delta = \frac{[\alpha_V]_{appr.} - [\alpha_V]_{ref.}}{[\alpha_V]_{ref.}}, \qquad (13)$$

where $[\alpha_V]_{appr.}$ is the approximate value given by Eq. 12 and $[\alpha_V]_{ref.}$ is the reference. The highly accurate reference values of the Voigt width, i.e. $[\alpha_V]_{ref.}$, can be obtained by solving Eq. 2 using the latest versions of MATLAB that supports the symbolic computing system with high number of significand precision.

Fig. 1 shows (a) the comparison of each width functions and (b) the logarithm of relative error for the width of Voigt profile, respectively, in the $R$, $D$ format. As we can see, in the Gaussian dominant region, i.e. $0 \leq y \leq 0.6993$(-1≤ $D$≤ -0.0870), the approximation Eq. 4 is highly accurate and provides accuracy better than $10^{-17}$. Numerical results illustrate that the calculation accuracy of Eq. 4 increases significantly with the decrease of $y$, which indicates that this expression is more suitable for the calculation of small $y$. In particular, the accuracy of Eq. 4 can reach the super accuracy of $10^{-34}$ (quadruple precision) in the domain $0 \leq y \leq 0.1974$(-1≤ $D$≤ -0.6167). Similarly, in the Lorentz dominant region, i.e. $y \geq 8.2507$(0.8167≤ $D$≤ 1), the approximation Eq. 8 is highly accurate and provides accuracy better than $10^{-17}$. Numerical results show that the calculation accuracy of Eq. 8 increases significantly with the increase of $y$, which indicates that this expression is more suitable for the calculation of large $y$. In particular, the accuracy of Eq. 8 can reach the super accuracy of $10^{-34}$ in the domain $y \geq 28.2058$(0.9427≤ $D$≤ 1). The Voigt width approximation is particularly important in the middle region, i.e. $0.6993 < y < 8.2507$(-0.0870 < $D$ < 0.8167), in which the width of the Gaussian profile is comparable to the width of the Lorentzian profile. Numerical results show that the approximation Eq. 11 based on the best uniform approximation of Chebyshev polynomial with 30 degree is highly accurate and provides accuracy better than $10^{-17}$ in the middle region. In summary, the approximation scheme Eq. 12 is highly accurate and provides accuracy better than $10^{-17}$ for arbitrary $\alpha_L/\alpha_G$ ratios, which indicates that the calculation error can be ignored for the general double precision calculation platform (16 bits).

4. Conclusion

A simple approximation scheme to describe the width of the Voigt profile as a function of the relative contributions of Gaussian and Lorentzian broadening is presented in this work. The numerical calculations suggest that the proposed approximation scheme can achieve super-accuracy (better than $10^{-17}$) calculation for Voigt profiles for arbitrary $\alpha_L/\alpha_G$ ratios. In particular, the accuracy reaches an astonishing $10^{-34}$ in the domain $0 \leq y \leq 0.1974 \cup y \geq 28.2058$($0 \leq \alpha_L/\alpha_G \leq 0.2371 \cup \alpha_L/\alpha_G \geq 33.8786$). Furthermore, the coefficients in Eq. 8 are all simple rational fractions, and the law behind this expression is also an interesting mathematical problem worth exploring.

**Acknowledgments**

We acknowledge financial support from the National Key Research and Development Program of China under Grant 2017YFB0603204.

**Appendix A**

In this part, we derive the expression of the ($r$+1)-th coefficient $p_r$ from the first $r$ coefficients $p_0, p_1, ..., p_{r-1}$ ($r$>=1) using the series analysis method.

$$\Gamma(y) = \sum_{n=0}^{r-1} p_n y^n + \sum_{n=r}^{\infty} p_n y^n. \tag{A.1}$$

*Step* 1: *Series expansion for the factors in Eq. 5 at y = 0.*

(1) Series expansion for $e^{\Gamma^2}$

The exponential function $e^{\Gamma^2}$ can be expanded into the following series at $y = 0$:

$$e^{\Gamma^2} = e^{p_0^2} + \sum_{n=1}^{\infty} \frac{a_n}{n!}(\Gamma - p_0)^n = e^{p_0^2} + \sum_{n=1}^{r-1} \frac{a_n}{n!}(\sum_{k=1}^{r-1} p_k y^k)^n + o(y^{r-1}), \tag{A.2}$$

where the coefficients $a_n$ is defined as:

$$a_n = \frac{d^n}{dt^n} e^{t^2} \Big|_{t=p_0} = e^{p_0^2} \sum_{k=\lceil \frac{n-1}{2} \rceil}^{n} \sum_{j=0}^{\lfloor k-(n-1)/2 \rfloor} \frac{(-1)^j p_0^{2k-n}(1-2j+2k-n)_n}{j!(-j+k)!}, \tag{A.3}$$

where $(.)_n$ is the Pochhammer symbol. According to the multinomial theorem, Eq. A.2 can be rewritten as follows in ascending order of the degree of $y$:

$$e^{\Gamma^2} = \sum_{k=0}^{r-1} d_k y^k + o(y^{r-1}), \tag{A.4}$$

where the coefficients $d_k$ is defined as:

$$d_k = \begin{cases} e^{p_0^2} & k = 0 \\ \sum_{n=1}^{k} \dfrac{\alpha_{n,k,k} a_n}{n!} & k \geq 1, \end{cases} \tag{A.5}$$

where the coefficients $\alpha_{n,k,j}$ is defined as:

$$\alpha_{n,k,j} = \sum_{\substack{n_1+n_2+\cdots+n_j=n \\ n_1+2n_2+\cdots+jn_j=k \\ n_1 \geq 0, \cdots, n_j \geq 0}} \dfrac{n!}{n_1! \cdots n_j!} p_1^{n_1} \cdots p_j^{n_j}. \tag{A.6}$$

(2) Series expansion for $e^{-\Gamma^2}$

In the same way, the exponential function $e^{-\Gamma^2}$ can be expanded into the following series at $y = 0$:

$$e^{-\Gamma^2} = \sum_{k=0}^{r-1} f_k y^k + (p_r b_1 + \sum_{n=2}^{r} \dfrac{\alpha_{n,r,r} b_n}{n!}) y^r + o(y^r), \tag{A.7}$$

where the coefficients $f_k$ is defined as:

$$f_k = \begin{cases} e^{-p_0^2} & k = 0 \\ \sum_{n=1}^{k} \dfrac{\alpha_{n,k,k} b_n}{n!} & k \geq 1, \end{cases} \tag{A.8}$$

where the coefficients $b_n$ is defined as:

$$b_n = \dfrac{d^n}{dx^n} e^{-x^2} \Big|_{x=p_0} = 2^n e^{-p_0^2} (-p_0)^n n! \sum_{k=0}^{\lfloor n/2 \rfloor} \dfrac{(-4)^{-k} p_0^{-2k}}{k!(-2k+n)!}. \tag{A.9}$$

(3) Series expansion for $\sin(2\Gamma y)$

In the same way, the sine function $\sin(2\Gamma y)$ can be expanded into the following series at $y = 0$:

$$\sin(2\Gamma y) = \sum_{m=1}^{r} g_m y^m + o(y^r), \tag{A.10}$$

where the coefficients $g_m$ is defined as:

$$g_m = \sum_{k=0}^{\lfloor \frac{m-1}{2} \rfloor} (-1)^k \dfrac{\beta_{1+2k,m,m-1}}{(1+2k)!}, \tag{A.11}$$

where the coefficients $\beta_{n,k,j}$ is defined as:

$$\beta_{n,k,j} = \sum_{\substack{n_0+n_1+\cdots+n_j=n \\ n_0+2n_1+\cdots+(j+1)n_j=k \\ n_0 \geq 0, \cdots, n_j \geq 0}} \dfrac{n!}{n_0! \cdots n_r!} 2^n p_0^{n_0} \cdots p_j^{n_j} \tag{A.12}$$

(4) Series expansion for $\cos(2\Gamma y)$

In the same way, the cosine function $\cos(2\Gamma y)$ can be expanded into the following series at $y = 0$:

$$\cos(2\Gamma y) = \sum_{m=0}^{r} h_m y^m + o(y^r), \tag{A.13}$$

where the coefficients $h_m$ is defined as:

$$h_m = \begin{cases} 1 & m = 0 \\ \sum_{k=1}^{\lfloor \frac{m}{2} \rfloor} \dfrac{(-1)^k \beta_{2k,m,m-1}}{(2k)!} & m \geq 1, \end{cases} \tag{A.14}$$

(5) Series expansion for $\mathrm{erfi}(\Gamma)$

In the same way, the imaginary error function $\mathrm{erfi}(\Gamma)$ can be expanded into the following series at $y = 0$:

$$\mathrm{erfi}(\Gamma) = \mathrm{erfi}(p_0) + \sum_{k=1}^{r-1} \sum_{n=1}^{k} \dfrac{\alpha_{n,k,k} e_n}{n!} y^k + o(y^{r-1}), \tag{A.15}$$

where the coefficients $e_n$ is defined as:

$$e_n = \frac{d^n \operatorname{erfi}(t)}{dt^n}\Big|_{t=p_0} = \frac{e^{p_0^2}}{\sqrt{\pi}} \sum_{k=\left\lceil \frac{n-2}{2} \right\rceil}^{n-1} \frac{2^{2+2k-n} p_0^{1+2k-n} (2+2k-n)_{2(-1-k+n)}}{(n-k-1)!}. \tag{A.16}$$

(6) Series expansion for erfc(y-iΓ)

The complementary error function erfc(y-iΓ) can be expanded into the following series at y = 0:

$$\operatorname{erfc}(y - i\Gamma) = 1 + i\operatorname{erfi}(\Gamma) - \frac{2}{\sqrt{\pi}} e^{\Gamma^2} \sum_{k=1}^{\infty} \sum_{j=\left\lceil \frac{k-1}{2} \right\rceil}^{k-1} \frac{(-i)^{1-k} 2^{1+2j-k} (2+2j-k)_{2(-1-j+k)} y^k \Gamma^{1+2j-k}}{k!(-1-j+k)!}, \tag{A.17}$$

and the third term in the above formula can be reduced to the following formula:

$$-\frac{2}{\sqrt{\pi}} e^{\Gamma^2} \sum_{k=1}^{\infty} \sum_{j=\left\lceil \frac{k-1}{2} \right\rceil}^{k-1} \frac{(-i)^{1-k} 2^{1+2j-k} (2+2j-k)_{2(-1-j+k)} y^k \Gamma^{1+2j-k}}{k!(-1-j+k)!} = e^{\Gamma^2} (\sum_{n=1}^{r} c_n y^n + o(y^r)), \tag{A.18}$$

where the coefficients $c_n$ is defined as:

$$c_n = -\frac{2}{\sqrt{\pi}} (\frac{(-1)^{n+1}+1}{2} \frac{1_{n-1} i^{1-n}}{n!((n-1)/2)!} + \sum_{k=1}^{n} \sum_{\substack{j=\left\lceil \frac{k-1}{2} \right\rceil \\ 1+2j-k\geq 1}}^{k-1} \chi_{1+2j-k,n-k,n-k} \frac{(-i)^{1-k} 2^{1+2j-k} (2+2j-k)_{2(-1-j+k)}}{k!(-1-j+k)!}), \tag{A.19}$$

where the coefficients $\gamma_{n,k,j}$ is defined as:

$$\gamma_{n,k,j} = \sum_{\substack{n_0+n_1+\cdots+n_j=n \\ n_1+2n_2+\cdots+jn_j=k \\ n_0\geq 0,\cdots,n_j\geq 0}} \frac{n!}{n_0!\cdots n_j!} p_0^{n_0} \cdots p_j^{n_j}. \tag{A.20}$$

Considering the expansion series of A.2 and A.18, the third term in Eq. A.17 can be reduced to the following formula:

$$-\frac{2}{\sqrt{\pi}} e^{\Gamma^2} \sum_{k=1}^{\infty} \sum_{j=\left\lceil \frac{k-1}{2} \right\rceil}^{k-1} \frac{(-i)^{1-k} 2^{1+2j-k} (2+2j-k)_{2(-1-j+k)} y^k \Gamma^{1+2j-k}}{k!(-1-j+k)!} = \sum_{m=1}^{r} \sum_{k=0}^{m-1} c_{m-k} d_k y^m + o(y^r). \tag{A.21}$$

By substituting Eq. A.15 and Eq. A.21 into Eq. A.17, the expressions of real part and imaginary part of the complementary error function erfc(y-iΓ) can be expanded into the following series:

$$\operatorname{Im}[\operatorname{erfc}(y - i\Gamma)] = \sum_{k=0}^{r-1} l_k y^k + o(y^{r-1}),$$

$$\operatorname{Re}[\operatorname{erfc}(y - i\Gamma)] = \sum_{k=0}^{r-1} q_k y^k + \sum_{k=0}^{r-1} \operatorname{Re}[c_{r-k} d_k] y^r + o(y^r), \tag{A.22}$$

respectively, where the coefficients $l_k$ and $q_k$ are defined as:

$$l_k = \begin{cases} \operatorname{erfi}(p_0) & k=0 \\ \sum_{n=1}^{k} \frac{\alpha_{n,k,k} e_n}{n!} + \sum_{n=0}^{k-1} \operatorname{Im}[c_{k-n} d_n] & k\geq 1, \end{cases} \tag{A.23}$$

$$q_k = \begin{cases} 1 & k=0 \\ \sum_{n=0}^{k-1} \operatorname{Re}[c_{k-n} d_n] & k\geq 1. \end{cases} \tag{A.24}$$

(7) Series expansion for erfc(y)

The error function erfc(y) can be expanded into the following series at y = 0:

$$\frac{1}{2}\operatorname{erfc}(y) = \sum_{n=0}^{r} s_n y^n + o(y^r), \tag{A.25}$$

where the coefficients $s_n$ is defined as:

$$s_n = \begin{cases} \frac{1}{2} & n=0 \\ -\frac{1+(-1)^{n+1}}{2} \frac{1}{\sqrt{\pi}} \frac{(-1)^{(n-1)/2}}{((n-1)/2)! n} & n\geq 1. \end{cases} \tag{A.26}$$

*Step 2: Calculate $p_r$ by the method of comparing coefficient.*

By substituting Eq. A.7, Eq. A.10, Eq. A.13, Eq. A.22 and Eq. A.25 into Eq. 5, the following expression is obtained:

$$[\sum_{k=0}^{r-1} f_k y^k][\sum_{m=1}^{r} g_m y^m][\sum_{k=0}^{r-1} l_k y^k] + [\sum_{m=0}^{r} h_m y^m][\sum_{k=0}^{r-1} f_k y^k + (p_r b_1 + \sum_{n=2}^{r} \frac{\alpha_{n,r,r} b_n}{n!}) y^r] \times$$
$$[\sum_{k=0}^{r-1} q_k y^k + \sum_{k=0}^{r-1} \text{Re}[c_{r-k} d_k] y^r] = \sum_{n=0}^{r} s_n y^n + o(y^r) \tag{A.27}$$

Therefore, a mathematical expression to determine the ($r$+1)-th coefficient $p_r$ is obtained by using the method of comparing coefficient:

$$\sum_{\substack{j+k+m=r \\ 0 \leq j \leq r-1, 0 \leq k \leq r-1, 1 \leq m \leq r}} f_j g_m l_k + \sum_{j=1}^{r} \sum_{k=0}^{r-j} q_{r-j-k} f_k h_j$$
$$+ h_0 [\sum_{k=1}^{r-1} q_{r-k} f_k + f_0 \sum_{j=0}^{r-1} \text{Re}[c_{r-j} d_j]] y^r + (p_r b_1 + \sum_{n=2}^{r} \alpha_{n,r,r} \frac{b_n}{n!}) q_0] = s_r. \tag{A.28}$$

The recurrence equation satisfied by the expression of the ($r$+1)-th coefficient $p_r$ can be obtained by simplifying the above expression:

$$p_r = \frac{e^{p_0^2}}{-2p_0} [s_r - q_r \exp(-p_0^2) - \sum_{\substack{j+k+m=r \\ 0 \leq j \leq r-1, 0 \leq k \leq r-1, 1 \leq m \leq r}} f_j g_m l_k - \sum_{j=1}^{r} \sum_{k=0}^{r-j} q_{r-j-k} f_k h_j$$
$$- \sum_{k=1}^{r-1} q_{r-k} f_k - \sum_{n=2}^{r} \frac{\alpha_{n,r,r} b_n}{n!}]. \tag{A.29}$$

Considering the initial value $p_0 = \sqrt{\ln 2}$, arbitrary term coefficients $p_n$ can be calculated by using formula Eq. A.29, and the first 31 coefficients $p_n$ with 32 significant digits are given in the Table A.1.

Table A.1 The first 31 coefficients $p_n$ in Eq. 4 with 32 significant digits.

| $n$ | $p_n$ | $n$ | $p_n$ |
|---|---|---|---|
| 0 | 0.83255461115769775635316464448952 | 16 | 2.5308903977393059634088084205148×10$^{-8}$ |
| 1 | 0.53254711842961210323020845059416 | 17 | -3.3104307709547517055285672959576×10$^{-8}$ |
| 2 | 0.13603423870145348659601346974136 | 18 | -1.1821070040002130133075915099552×10$^{-8}$ |
| 3 | -6.3839925995348583105863651935208×10$^{-3}$ | 19 | 5.0020607880755762331999675884955×10$^{-9}$ |
| 4 | -7.5882994178697868047017954181619×10$^{-3}$ | 20 | 3.2040951850692659104678394048668×10$^{-9}$ |
| 5 | 7.5685451134845100193553849814044×10$^{-4}$ | 21 | -4.9276721508012916216290609360574×10$^{-10}$ |
| 6 | 6.4174309726033170181322853645455×10$^{-4}$ | 22 | -7.1352246104725448681836423474852×10$^{-10}$ |
| 7 | -1.0278614365257442345642575963235×10$^{-5}$ | 23 | -3.2407999521382539130974667691197×10$^{-11}$ |
| 8 | -6.6864392638387619203117167133824×10$^{-5}$ | 24 | 1.4010883014405512366881008147675×10$^{-10}$ |
| 9 | -1.8800729899141457354675112660009×10$^{-5}$ | 25 | 3.3772678382804066494130831543588×10$^{-11}$ |
| 10 | 9.3901358253570724565409358708571×10$^{-6}$ | 26 | -2.3680267709485323621904030934022×10$^{-11}$ |
| 11 | 5.4149990265667553408636905696295×10$^{-6}$ | 27 | -1.1462686830778835681784218983719×10$^{-11}$ |
| 12 | -1.2862976252461744893956942201673×10$^{-6}$ | 28 | 3.0039670445166124668988923778107×10$^{-12}$ |
| 13 | -1.0759168918380548822306060203341×10$^{-6}$ | 29 | 2.9478889620399924642669859987364×10$^{-12}$ |
| 14 | 7.8733635964790862989086501951507×10$^{-8}$ | 30 | -9.7467645599626148566298439388065×10$^{-14}$ |
| 15 | 1.9255725519174188542320412973488×10$^{-7}$ | | |

**Appendix B**

The relationship between $\Gamma'^2$ and $\eta$ can be written in the form of Taylor series as following:

$$\Gamma'^2(\eta) = \sum_{n=0}^{\infty} T_n \eta^n, \tag{B.1}$$

where $T_0 = 1$ and the coefficients $T_n$ ($n > 1$) can be estimated by the following recurrence relations:

$$T_n = \lim_{\eta \to 0^+} \frac{\Gamma'^2(\eta) - \sum_{m=0}^{n-1} T_m \eta^m}{\eta^n} \approx \frac{\Gamma'^2(\varepsilon) - \sum_{m=0}^{n-1} T_m \varepsilon^m}{\varepsilon^n} \triangleq T_n^* \ (0 < \varepsilon \ll 1). \tag{B.2}$$

In this work, parameter $\varepsilon$ in above equation is set to $1 \times 10^{-20}$ and the benchmark value of $\Gamma'(\varepsilon)$ is obtained according to Eq. 2 by using the latest versions of MATLAB that supports the symbolic computing system with high number of significand precision (1000 bits). The coefficients $T_n$ and $T_n^*$ of the first 31 terms are given in the Table B.1. Numerical calculation results suggest that the $T_n^*$ value can be rewritten as the sum of a rational fraction and a fairly small remainder. Fortunately, the reduction in accuracy due to the elimination of the remainder is negligible.

Table B.1 The first 31 coefficients $T_n$ in Eq. 8.

| n | $T_n^*$ | $T_n$ | n | $T_n^*$ | $T_n$ |
|---|---|---|---|---|---|
| 0 | / | 1 | 16 | $\approx -25390179/2^{11}+1.1\times10^{-35}$ | $-25390179/2^{11}$ |
| 1 | $\approx 0+1.5\times10^{-20}$ | 0 | 17 | $\approx 0+1.1\times10^{-15}$ | 0 |
| 2 | $\approx 3/2-7.5\times10^{-41}$ | 3/2 | 18 | $\approx 446848569/2^{12}-1.1\times10^{-34}$ | $446848569/2^{12}$ |
| 3 | $\approx 0-7.5\times10^{-21}$ | 0 | 19 | $\approx 0+-1.1\times10^{-14}$ | 0 |
| 4 | $\approx -3/2^2+1.9\times10^{-40}$ | $-3/2^2$ | 20 | $\approx -1089694161/2^{10}+1.1\times10^{-33}$ | $-1089694161/2^{10}$ |
| 5 | $\approx 0+1.9\times10^{-20}$ | 0 | 21 | $\approx 0+1.1\times10^{-13}$ | 0 |
| 6 | $\approx 15/2^3-7.6\times10^{-40}$ | $15/2^3$ | 22 | $\approx 46704949839/2^{12}-1.3\times10^{-32}$ | $46704949839/2^{12}$ |
| 7 | $\approx 0-7.6\times10^{-20}$ | 0 | 23 | $\approx 0-1.3\times10^{-12}$ | 0 |
| 8 | $\approx -243/2^5+3.9\times10^{-39}$ | $-243/2^5$ | 24 | $\approx -8735832539883/2^{16}+1.7\times10^{-31}$ | $-8735832539883/2^{16}$ |
| 9 | $\approx 0+3.9\times10^{-19}$ | 0 | 25 | $\approx 0+1.7\times10^{-11}$ | 0 |
| 10 | $\approx 2493/2^6-2.3\times10^{-38}$ | $2493/2^6$ | 26 | $\approx 221377058104455/2^{17}-2.3\times10^{-30}$ | $221377058104455/2^{17}$ |
| 11 | $\approx 0-2.3\times10^{-18}$ | 0 | 27 | $\approx 0-2.3\times10^{-10}$ | 0 |
| 12 | $\approx -927/2^2+1.6\times10^{-37}$ | $-927/2^2$ | 28 | $\approx -6044700753428715/2^{18}+3.4\times10^{-29}$ | $-6044700753428715/2^{18}$ |
| 13 | $\approx 0+1.6\times10^{-17}$ | 0 | 29 | $\approx 0+3.4\times10^{-9}$ | 0 |
| 14 | $\approx 405783/2^8-1.2\times10^{-36}$ | $405783/2^8$ | 30 | $\approx 353911508743725891/2^{20}+2.9\times10^{-6}$ | $353911508743725891/2^{20}$ |
| 15 | $\approx 0-1.2\times10^{-16}$ | 0 | | | |

**Appendix C**

Table C.1 The first 31 coefficients $u_n$ in Eq. 11 with 32 significant digits.

| n | $u_n$ | n | $u_n$ |
|---|---|---|---|
| 0 | 0.81879767981374096480451126966969 | 16 | -11.608186550060767559947858982236 |
| 1 | -0.087358831239253690600565585478191 | 17 | 98.530866614251729915080851559823 |
| 2 | 0.16111263881308988360982026625923 | 18 | -520.58001078415212154632736105054 |
| 3 | 0.10352476879958392716101379868109 | 19 | 1996.0992356052342655084175033613 |
| 4 | 0.04470194137424132479415258752939 | 20 | -5861.8655902675083091764869931816 |
| 5 | -0.0014922440275783965022004229833442 | 21 | 13508.475417035315538373423012345 |
| 6 | -0.025999766558392062049748550766996 | 22 | -24679.618517222644663493302537093 |
| 7 | -0.027433278283219905509735236616617 | 23 | 35800.136359576107078418696532107 |
| 8 | -0.012324451041403454228824532372558 | 24 | -41003.133466095685472430284467431 |
| 9 | 0.0076580003679061144826236693340858 | 25 | 36602.695793508138571786897415062 |
| 10 | 0.020609356479185309053858762785961 | 26 | -24908.669183448932665943004856528 |
| 11 | 0.019910337726501870836014131102016 | 27 | 12466.933666792486921786288 6569 |
| 12 | 0.0077590409364772777791275317929576 | 28 | -4321.2090018088841034237110866818 |
| 13 | -0.047206722791611042489905888231109 | 29 | 925.71414182435000624708620108582 |
| 14 | 0.16487160565916573888004662456667 | 30 | -92.274791451466791656684619 21968 |
| 15 | 0.22613547743220062910838574670492 | | |